\documentclass[times,11pt,a4paper]{article}
\usepackage[english]{babel}                
\usepackage{graphicx}
\usepackage{amsmath,amssymb}  
\usepackage[utf8]{inputenc}
\usepackage{amsfonts}
\usepackage{natbib}
\bibpunct{(}{)}{;}{}{}{,}
\usepackage{booktabs}
\usepackage{epstopdf}
\usepackage{multicol}
\usepackage{makecell}
\usepackage{array}
\usepackage[dvipsnames]{xcolor}
\usepackage[nottoc,notlot,notlof]{tocbibind} 
\usepackage[top=1.5in, bottom=1.5in, left=1in, right=1in]{geometry}
\usepackage{color,colortbl} 
\usepackage[toc,page]{appendix} 
\usepackage{caption} 
\addto\captionsenglish{}
\usepackage{gensymb}
\usepackage{bm}
\usepackage{textcomp}
\usepackage{booktabs}

\usepackage{floatrow}
\usepackage{subfig}
\usepackage{wrapfig}
\usepackage{hhline}

\begin{document}

\newcommand{\degr}     {\textdegree\:} 
\newcommand{\degrs}    {\textdegree S\:}
\newcommand{\degrn}    {\textdegree N\:}
\newcommand{\degre}    {\textdegree E\:}
\newcommand{\degrw}    {\textdegree W\:}
\newcommand{\degrc}    {\textdegree C\:}
\newcommand{\minusone} {$^{-1}$}
\newcommand{\ordinal}  {$^{th}$\:}
\newcommand{\erase}    {\bgroup\markoverwith{\textcolor{red}{\rule[.5ex]{2pt}{1.5pt}}}\ULon}
\newcommand{\new}   {\textcolor{red}}
\newcommand{\rpm}{\raisebox{.2ex}{$\scriptstyle\pm$}}

\begin{center}
\textbf{\Large Synoptic circulation patterns during temperature extremes in southeastern Europe} 
\end{center}
\vspace{8mm}

\textbf{Fragkoulidis G.$^{\text{\normalfont 1*}}$, Wirth V.$^{\text{\normalfont 1}}$}\vspace{3mm}

\small{1 Institute for Atmospheric Physics, Johannes Gutenberg University, Mainz, Germany}\vspace{3mm}

\small{$\text{*}$ corresponding author e-mail: gfragkou@uni-mainz.de} 

\vspace{4mm}

\begin{abstract}
This study investigates the synoptic circulation patterns associated with temperature extremes in southeastern (SE) Europe. Using ERA-Interim reanalysis data we report on the typical patterns that characterize the middle and upper tropospheric flow a few days before and during hot and cold extremes. The analysis is done separately for each season, while a further distinction between short-lived and persistent extremes reveals differences and similarities in the associated circulation, as inferred by the spatiotemporal evolution of Rossby wave packets (RWPs). Finally, the performance of ECMWF deterministic forecasts is evaluated for persistent cold and hot extremes during winter and summer respectively. Overall, this work suggests that a correct representation of the RWP evolution is crucial in determining the magnitude and persistence of temperature extremes in SE Europe.
\end{abstract}

\section{Introduction} \label{intro}
A better understanding of the physical processes that lead to weather extremes is essential for various reasons. From a weather forecast perspective, it proves beneficial in the challenges of identifying the predictability limits and interpreting forecast biases or ``busts" associated with these events \citep{Rodwell2013}. From a climate perspective, investigating further the relevant processes will help in formulating hypotheses and evaluating model projections on the weather extremes characteristics of futures decades \citep{Shepherd2014}. In this regard, recent studies have examined the possibility of far-upstream precursors to weather extremes by tracking transient Rossby Wave Packets (RWPs) in the upper-tropospheric flow \citep{Martius2008,Fragkoulidis2018,Wirth2018}.

This study focuses on the role of the middle and upper tropospheric circulation during abnormally hot and cold spells in southeastern (SE) Europe. This region is particularly interesting in terms of large-scale dynamics as it lies to the southeast of the climatological North Atlantic jet exit (where Rossby waves are typically in their mature nonlinear phase and frequently wave breaking is observed) and slightly northward of the subtropical jet \citep{Sprenger2017}. Using anomaly composites of reanalysis and forecast data we report on typical circulation patterns and forecast errors in the daily evolution of short-lived and persistent temperature extremes of both signs.

\section{Data and Methodology} \label{data}
\subsection{Data}

Reanalysis data for meridional wind $v$ at 300~hPa, geopotential height $Z$ at 500~hPa, and temperature $T$ at 850~hPa spanning a 38-year period (1979-2016) are retrieved on a 2\degr$\times$2\degr horizontal grid with 6-hourly temporal resolution from the ERA-Interim reanalysis project \citep{Dee2011}. In addition, we use deterministic forecasts that were produced from the same ECMWF model (Integrated Forecasting System (IFS) model cycle 31r2) with the ERA-Interim analyses as initial conditions \citep{Berrisford2009}. Namely, 10-day forecasts at 12-hourly steps of $v$ at 300~hPa and $T$ at 850~hPa, issued daily at 1200~UTC, are retrieved on a 2\degr$\times$2\degr horizontal grid. 

\subsection{Methodology}

\begin{wrapfigure}[13]{r}{0.24\textwidth}
\includegraphics[trim = 0cm 0cm 0cm 0cm,clip,width=1.0\textwidth]{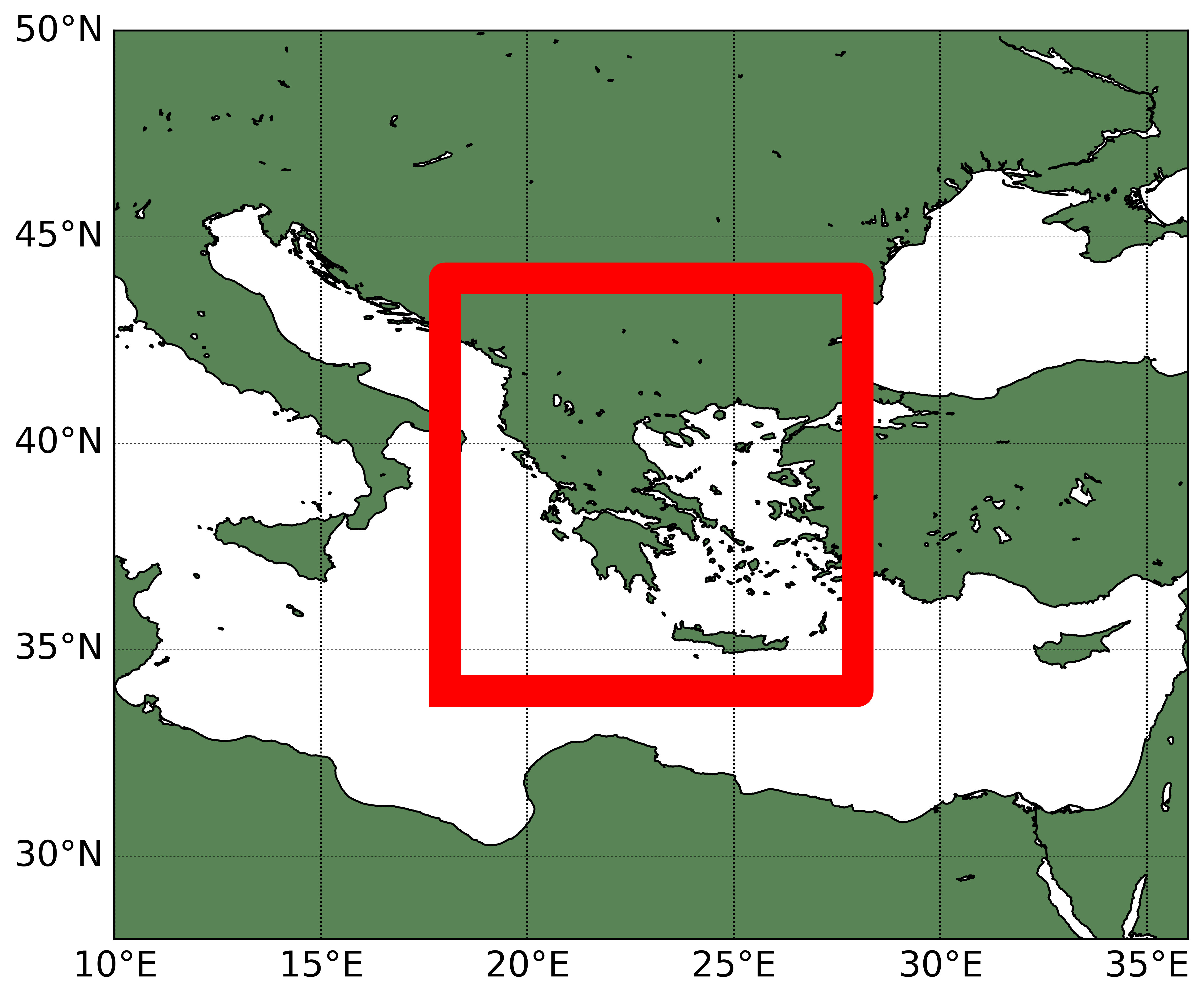}
\caption{Region of $T'$ averaging in the extremes selection.}\label{fig1}
\end{wrapfigure}

For all the retrieved data we calculate the daily mean anomalies ($v'$, $Z'$, $T'$) from a climatological annual cycle as in \citet{Fragkoulidis2018}. Since we are primarily interested in temperature extremes of large spatial extent in SE Europe, the following methodology to define hot and cold extreme events is used. We first detrend the field of 850~hPa $T'$ by subtracting the 1979--2016 linear trend at each grid point (the detrended field is only used for the selection of the temperature extreme events in this study). The detrended 850~hPa $T'$ is averaged over the region [34-44\degrn, 18-28\degre] (Fig.~\ref{fig1}) with a cosine latitude weighting. Days with an area-averaged $T'$ greater (lower) than the 95\ordinal (5\ordinal) percentile of the respective season (DJF, MAM, JJA, SON) are regarded hot (cold) extremes. Episodes of 1-2 extreme days constitute a \emph{short-lived} temperature extreme event. Episodes of 3 or more consecutive extreme days constitute a \emph{persistent} temperature extreme event. In few cases where the onset of two subsequent extreme events occurred within 4 days or less, we did not regard the second one to be distinct and independent from the first one and we discard it from the event list.

Due to the limited size of the event samples, composite fields in the following analyses are computed using the mean of the interquartile range of the sample, which restricts the effect of outliers and renders the statistics more robust. In addition, the statistical significance of the $v'$, $Z'$ and $T'$ composites is assessed with a Monte Carlo technique \citep[e.g.][]{Martius2008}. Namely, the values at each grid point are deemed significant at the $\alpha$=0.10 level, if they belong to either 5\% tail of a distribution created by reconstructing the composites 300 times using random selections of dates.

\section{Results} \label{chap1}

\setlength{\arrayrulewidth}{0.3mm}
\renewcommand{\arraystretch}{1.1}
\newcolumntype{M}[1]{>{\centering\arraybackslash}m{#1}}
\newcolumntype{P}[1]{>{\centering\arraybackslash}p{#1}}
\newcolumntype{s}{>{\columncolor{lightgray}} M{1.5cm}}
\arrayrulecolor{black}
 
\begin{wraptable}{r}{0.52\textwidth}
\centering 
\begin{tabular}{M{1.0cm}|M{1.3cm}M{1.3cm}|M{1.3cm}M{1.3cm}} \hline  
\multicolumn{1}{M{0.9cm}|}{} & \multicolumn{2}{M{3.0cm}|}{\cellcolor{Goldenrod}Hot extremes} & \multicolumn{2}{M{3.0cm}}{\cellcolor{SkyBlue}Cold extremes}\\ \hline
Season & Events & Duration & Events & Duration\\ \hline
DJF & 95 (16) & 1.8 (4.1) & 86 (21) & 2.0 (3.7) \\ 
MAM & 82 (26) & 2.1 (3.8) & 62 (30) & 2.8 (4.4) \\ 
JJA & 69 (26) & 2.5 (4.5) & 76 (25) & 2.3 (4.0) \\ 
SON & 70 (24) & 2.5 (4.4) & 80 (26) & 2.2 (3.6) \\ \hline
\end{tabular}
\caption{Number and mean duration (in days) of hot and cold extremes. Values in the parenthesis correspond to persistent temperature extremes.}
\label{my-label}
\end{wraptable} 
 
The resulting number and mean duration of distinct hot and cold extremes in all four seasons is shown in Table~1. The winter season is characterized by the shortest duration in extremes (1.8 days for hot and 2.0 for cold). The percentage of persistent extremes maximizes in summer for hot extremes (26/69, with 4.5 days mean duration) and spring for cold extremes (30/62, with 4.4 days mean duration). Finally, with the exception of spring, hot extremes in all seasons tend to last longer than cold extremes.  

Composites of 850~hPa $T'$, 500~hPa $Z'$ and 300~hPa $v'$ during JJA hot extremes (69 distinct events) are shown in Fig.~2a,c. In the $T'$ composite over the onset days (Fig.~2a), a dipole structure is evident with warm anomalies maximized over SE Europe and cold anomalies over Western Europe, both having a SW-NE orientation. In Fig.~2c, $v'$ and $Z'$ composites are shown for selected days before and after the onset of the events, revealing a distinct pattern of RWP propagation. More specifically, a trough-ridge sequence moves eastward at slow phase velocities, while the encompassing RWP as a whole propagates faster indicating the transfer of energy and generation of new disturbances downstream. Eventually, the upper-tropospheric flow over SE Europe is deflected northward forming a ridge with anticyclonic vorticity. As a consequence, we get a favorable environment for extreme temperatures since warm and dry air is advected from N. Africa. Previous studies \citep[e.g.][]{Founda2009} have shown that this works in conjuction with adiabatic compressional heating due to enhanced subsidence (which during summer months is anyway present as a climatic feature of the region) and anomalous radiative forcing to result in extreme temperatures.

\begin{figure}[!ht]
\includegraphics[trim = 0mm 0cm 0mm 0cm,clip,width=0.95\textwidth]{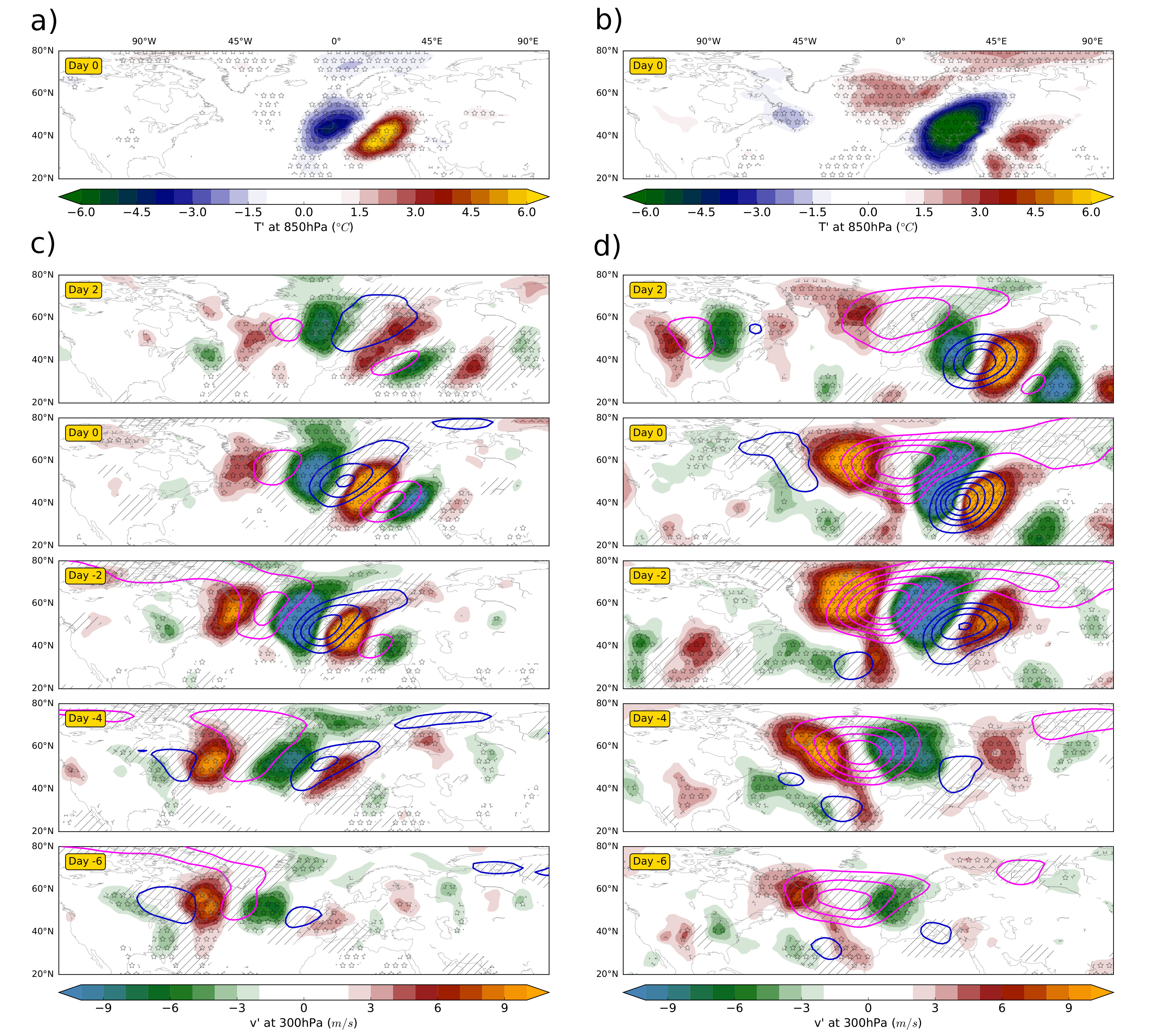}
\caption{a) Composite of 850~hPa $T'$ (color shading) at the onset of JJA hot extremes. Stars denote statistically significant values of $T'$ at $\alpha$=0.10. b) Same as a) but for DJF cold extremes. c) Evolution of 300~hPa $v'$ (color shading) and 500~hPa $Z'$ (blue/magenta contours indicate negative/positive anomalies at \rpm 3, \rpm 6, \rpm 9, ... gpdam) composites from Day -6 to Day 2 of JJA hot extremes. Stars (hatching) denote statistically significant values of $v'$ ($Z'$) at $\alpha$=0.10. d) Same as c) but for DJF cold extremes.}\label{fig2}
\end{figure}

Apart from the opposite sign in the anomaly composites during DJF cold extremes (86 distinct events), we get a rather different circulation pattern evolution (Fig.~2b,d). The areas affected by cold temperatures have a higher spatial extent and are surrounded by smaller centres of positive $T'$ (Fig.~2b). A distinct blocking anticyclone is a prevalent feature of the North Atlantic circulation several days before the onset of the events; a typical flow regime for cold extremes in Europe \citep{sillmann2009present}. At Day -2, an elongated trough has formed at the eastern flank of the block, transporting continental cold air masses from the northeast towards SE Europe. This pattern persists and two days later we have the onset of the cold extreme. Afterwards, the North Atlantic block starts to weaken and the RWP propagates into Asia arching toward the Tropics.

\begin{figure}[!ht]
\includegraphics[trim = 0cm 0cm 0cm 0cm,clip,width=0.95\textwidth]{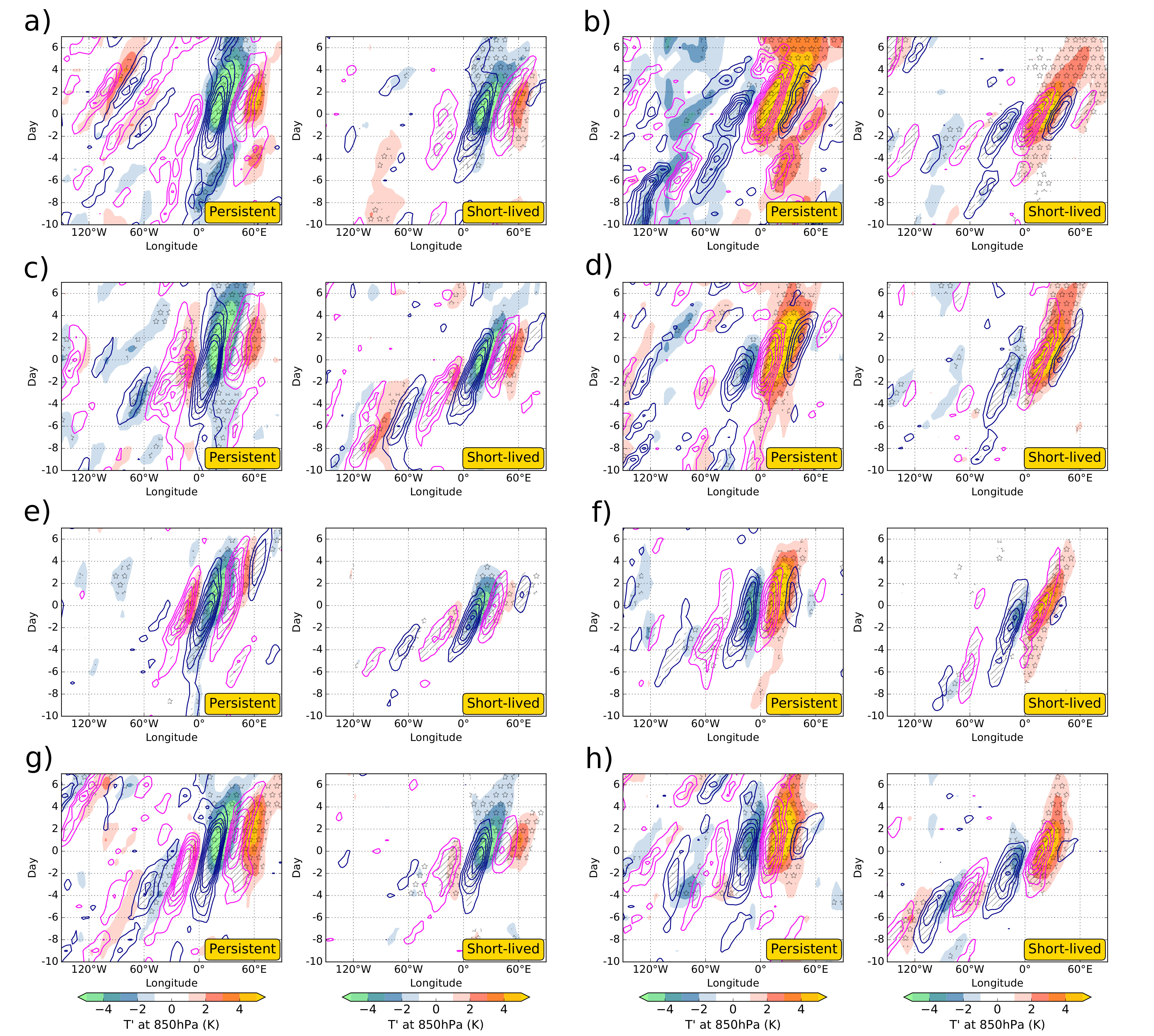}
\caption{a) Hovmöller composites of 850~hPa $T'$ (color shading) and 300~hPa $v'$ (blue/magenta contours indicate negative/positive anomalies at \rpm 3, \rpm 5, \rpm 7, ... m/s) during DJF persistent (left) and short-term (right) cold extremes. Stars (hatching) denote statistically significant values of $T'$ ($v'$) at $\alpha$=0.10. b) Same as a) but for hot extremes. c), d) Same for MAM. e), f) Same for JJA. g), h) Same for SON.}\label{fig3}
\end{figure}

We now consider short-lived and persistent temperature extremes  separately, in order to explore differences in the typical circulation associated with them. To this end, we show composites of 850~hPa $T'$ and 300~hPa $v'$ compressed in Hovmöller diagrams for all seasons and both types of extremes (Fig.~\ref{fig3}). $T'$ has been averaged over the 34-44\degrn latitudinal band, while $v'$ has been averaged over the wider 34-64\degrn latitudinal band, in order to capture the main features of the large-scale midlatitude flow. In composites of persistent extremes our sample size decreases considerably (Table~1) at the expense of statistical significance, but indications of qualitative differences from short-lived extremes can still arise. Some key observations on the various extreme types include:
\begin{enumerate}
\item[(i)] Patterns of $T'$ for the persistent extremes appear more elongated (in the time direction, as expected), wider (spanning more longitudes) and more stagnant (forming an almost right angle with the longitudes axis) than the ones for short-lived extremes.

\item[(ii)] In many cases (e.g. MAM short-lived cold extremes, DJF persistent hot extremes, SON short-lived hot extremes), the succession of positive (southerly) and negative (northerly) $v'$ values indicates a far-upstream RWP signal several days before the extreme onset. In the cases where an RWP is not traceable several days in advance, either an organized group of Rossby wave activity was not present (e.g. block cases in DJF cold extremes) or the RWPs evolve in varying pathways leading to partial destructive interference and a reduced wave signal from the sample. 

\item[(iii)] As is well known, the Hovmöller diagram can provide an estimation of the phase and the group velocity associated with an RWP \citep{Fragkoulidis2018}. In all cases, the group velocity is greater than the phase velocity of individual troughs-ridges. This so-called downstream development phenomenon is particularly pronounced for e.g. DJF short-lived hot extremes and JJA short-lived cold extremes and less so for e.g. JJA short-lived hot extremes and MAM short-lived extremes.

\item[(iv)] Persistent extremes are associated with quasi-stationary $v'$ patterns, whereas higher phase velocities are observed in short-lived extremes. This manifests the decisive role of upper-tropospheric circulation in the duration of temperature extremes \citep{Kysely2008}.

\item[(v)] The zonal gradient of $v'$ above SE Europe is greater for cold than hot extremes (also evident in Fig.~\ref{fig2}). Associated with that, hot extremes tend to occur close to the centre of the overlying ridge, while cold extremes occur closer to the area of negative $v'$ values (northerlies) of the trough aloft.
\end{enumerate}

Finally, it's worth verifying the performance of deterministic forecasts of these temperature extremes. The availability of a 38-year dataset with forecasts produced with the same model (section~2) allows us to have a consistent comparison with the corresponding reanalysis of selected events. Our verification involves forecasts issued 3 and 5 days prior to DJF cold and JJA hot persistent extremes (21 and 26 events respectively). Fig.~\ref{fig4} shows Hovmöller composites of $T'$ and $v'$ (as in Fig.~\ref{fig3}) for these forecasts. These are complemented by the forecast-minus-reanalysis Hovmöller composites, that serve to reveal systematic errors in the prediction of persistent extremes.   

\begin{figure}[!h]
\includegraphics[trim = 0cm 0cm 0cm 0cm,clip,width=0.95\textwidth]{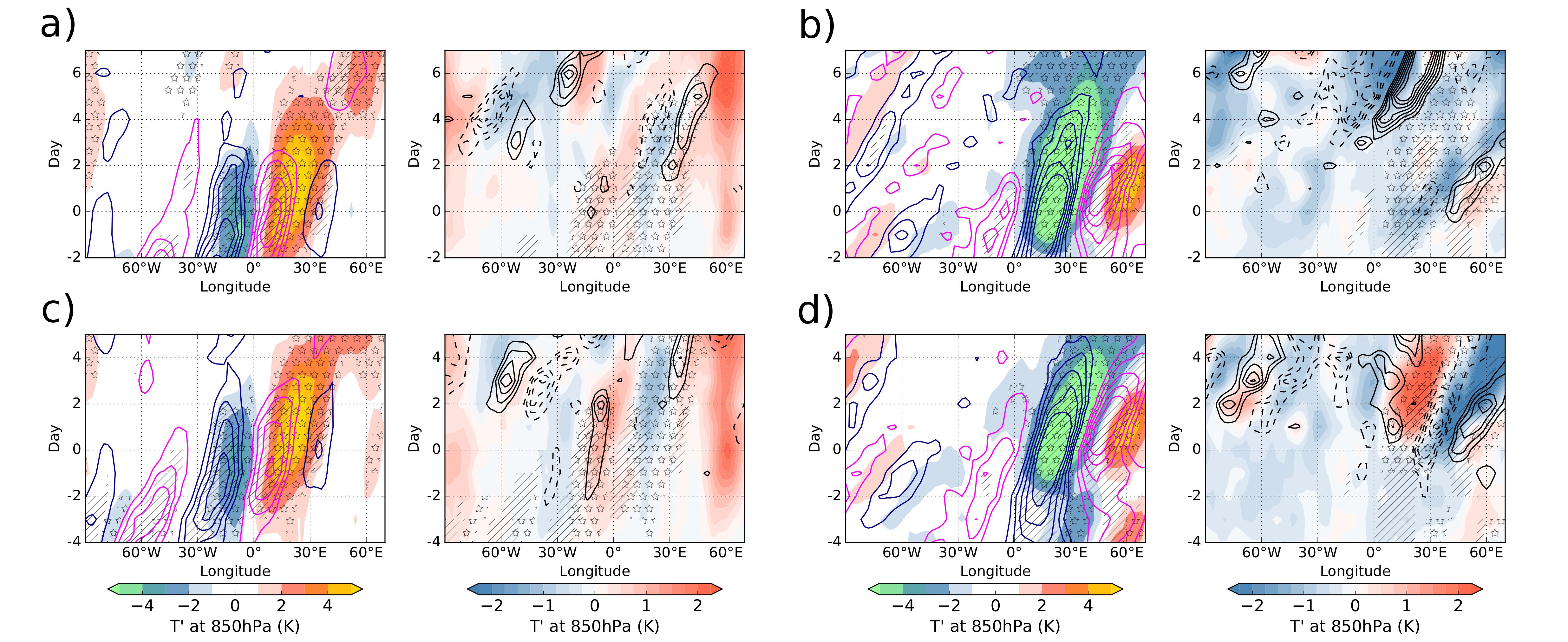}
\caption{a) Left: Hovmöller composites of 850~hPa $T'$ (color shading) and 300~hPa $v'$ forecasts (blue/magenta contours indicate negative/positive anomalies at \rpm 3, \rpm 5, \rpm 7, ... m/s), issued 3 days prior to JJA persistent hot extremes. Stars (hatching) denote statistically significant values of $T'$ ($v'$) at $\alpha$=0.10. Right: Deviation of the forecasted 850~hPa $T'$ (color shading) and 300~hPa $v'$ (solid/dashed contours indicate positive/negative anomalies at \rpm 2, \rpm 3, \rpm 4, ... m/s) from the respective reanalysis fields (Fig.~3f left). Statistically significant deviations are assumed where both the forecast and reanalysis fields are significant at $\alpha$=0.10. b) Same as a) but for DJF persistent cold extremes. c),d) Same as a),b) but for forecasts issued 5 days prior to the extremes.}\label{fig4}
\end{figure}

Forecasts issued 3 and 5 days prior to JJA persistent hot extremes show an underestimation of $T'$ above our region of interest (18-28\degre) and an overestimation to the east and west of that. This pattern can arise due to positional (some extremes are predicted to occur to the east and others to the west of the eventually affected region) and/or magnitude errors. Forecasts issued 3 days prior to DJF persistent cold extremes reveal an overestimation in both magnitude and duration. In 5-day forecasts, an eastward shift of the cold anomalies is predicted. In contrast to hot extremes, now an error is also evident in the $v'$ field with an eastward shift of the trough that can in principle explain the $T'$ error pattern.

\section{Conclusions}\label{s:discconc}
In this study, we investigated the synoptic circulation associated with the most pronounced temperature extremes that affected SE Europe in the last 38 years. Although the amount of such high-impact events is limited and robust results are not feasible, qualitative differences and similarities between the synoptic circulation imprints of hot and cold extremes across all seasons were inferred. In addition, distinguishing between short-lived and persistent extremes revealed differences in the spatiotemporal evolution of the upper-tropospheric flow. Far-upstream precursors were evident for several types of temperature extremes, indicating a recurrent in-phase RWP propagation, while in other cases the less clear signal in the days preceding the extremes suggests a larger variability with no preferential RWP pathway. 

Future studies on medium-range forecasts of temperature extremes should investigate further the role of upper-tropospheric dynamical features and how model limitations lead to systematic errors in the Rossby wave structure \citep{Gray2014a}. Moreover, an improved understanding on the interplay between synoptic circulation patterns, radiative transfer and boundary layer processes will help in interpreting the observed increase in frequency and severity of persistent temperature extremes \citep{Kysely2008}.

\section*{Acknowledgements}
We thank the ECMWF for providing access to the data used in this study. The research leading to these results has been done within the Transregional Collaborative Research Center SFB/TRR 165 ``Waves to Weather” funded by the German Research Foundation (DFG).

\bibliographystyle{wileyqj}
\bibliography{references}

\end{document}